\documentclass[sigconf]{acmart}

\def\BibTeX{{\rm B\kern-.05em{\sc i\kern-.025em b}\kern-.08emT\kern-.1667em\lower.7ex\hbox{E}\kern-.125emX}}
    
\copyrightyear{2021}
\acmYear{2021}
\setcopyright{iw3c2w3}
\acmConference[WWW '21]{Proceedings of the Web Conference 2021}{April 19--23, 2021}{Ljubljana, Slovenia}
\acmBooktitle{Proceedings of the Web Conference 2021 (WWW '21), April 19--23, 2021,
Ljubljana, Slovenia}

\acmPrice{}
\acmDOI{10.1145/3442381.3450009}
\acmISBN{978-1-4503-8312-7/21/04}

\usepackage{multirow}
\usepackage{diagbox}
\newcommand{\tabincell}[ 2 ]{\begin{tabular}{@{}#1@{}}#2\end{tabular}}

\begin{document}

\title{A Linguistic Study on Relevance Modeling in Information Retrieval}

\author{Yixing Fan, Jiafeng Guo, Xinyu Ma, Ruqing Zhang, Yanyan Lan, and Xueqi Cheng}
\email{{fanyixing, guojiafeng, maxinyu18s, zhangruqing, lanyanyan, cxq}@ict.ac.cn}
\affiliation{%
  \institution{University of Chinese Academy of Sciences, Beijing, China\\
  CAS Key Lab of Network Data Science and Technology, Institute of Computing Technology,\\ Chinese Academy of Sciences, Beijing, China\\}
}

\begin{abstract}

Relevance plays a central role in information retrieval (IR), which has received extensive studies starting from the 20th century. The definition and the modeling of relevance has always been critical challenges in both information science and computer science research areas. Along with the debate and exploration on relevance, IR has already become a core task in many real-world applications, such as Web search engines, question answering systems, conversational bots, and so on. While relevance acts as a unified concept in all these retrieval tasks, the inherent definitions are quite different due to the heterogeneity of these tasks. This raises a question to us: Do these different forms of relevance really lead to different modeling focuses? To answer this question, in this work, we conduct an empirical study on relevance modeling in three representative IR tasks, i.e., document retrieval, answer retrieval, and response retrieval. Specifically, we attempt to study the following two questions: 1) Does relevance modeling in these tasks really show differences in terms of natural language understanding (NLU)? We employ 16 linguistic tasks to probe a unified retrieval model over these three retrieval tasks to answer this question. 2) If there do exist differences, how can we leverage the findings to enhance the relevance modeling? 
We proposed three intervention methods to investigate how to leverage different modeling focuses of relevance to improve these IR tasks. We believe the way we study the problem as well as our findings would be beneficial to the IR community.

\end{abstract}

%
%

\ccsdesc[500]{Computer systems organization~Embedded systems}
\ccsdesc[300]{Computer systems organization~Redundancy}
\ccsdesc{Computer systems organization~Robotics}
\ccsdesc[100]{Networks~Network reliability}

%
\keywords{relevance modeling, information retrieval}

%
\maketitle

\section{introduction}
Information retrieval (IR) has already became a ubiquitous activity in our daily life. People rely on IR systems to obtain information that is relevant to their needs. Relevance, which denotes how well a retrieved document meets the information need of a user, plays a central role in IR. In fact, all the retrieval models in IR systems are trying to approximate the relevance from the perspective of users. However, the concept of relevance, likes all the other human notions, is an open and vague subject~\cite{saracevic2015relevance}.

It has been a long-standing challenge to understand and model relevance in two major research communities, i.e., information science community and computer science community.
On one hand, researchers from information science community studied the definition of relevance concept since 1950s~\cite{hillman1964notion, saracevic1975relevance, mizzaro1997relevance}. They tried to uncover the aspects of the relevance based on the data collected from tests or questionnaires. 
On the other hand, researchers from computer science community mainly focused on the modeling/computation of relevance since the mid-1960s~\cite{maron1960relevance}. A large number of models have been proposed to evaluate the relevance degree of a document with respect to users' information needs~\cite{li2014semantic, harman2019information}. These models have evolved from shallow to deep understanding of the document and the information need, which are often based on heuristically designed features or functions.
However, there has been few studies to take the relevance definition into account in designing relevance models.




Along with the debate and exploration on relevance, IR has been widely applied and become a core task in many real-world applications, such as Web search engines, question answering systems, conversational bots, and so on. In Web search engines, the IR task is to rank a list of documents according to their relevance to a given user query. In question answering systems, the IR task is to retrieve a few relevant answers from the archived answer pool with respect to a user's question. In conversational bots, the IR task is to find the relevant response from existing human-generated conversation repository as the reply to the input utterance. 
Without loss of generality, relevance acts as a unified concept in all these IR tasks. However, we may find subtle differences on the  definition of the relevance concept among these tasks. 
For example, the relevant documents in Web search often means topical relevance to the search query~\cite{li2014semantic}. The relevant answers in question answering need to correctly address the question~\cite{bajaj2016ms}. Finally, the relevant responses in conversation actually refer to some kind of correspondence with respect to the input utterance~\cite{lowe2015ubuntu}. In summary, the inherent definitions of relevance actually are quite different due to the heterogeneity of different IR tasks~\cite{guo2019deep}.

The above observations naturally raise a question to us:
Do different forms of relevance in these IR tasks really lead to different modeling focuses? To answer this question, in this paper, we conduct an empirical study to investigate the relevance modeling in three representative IR tasks, namely document retrieval, answer retrieval, and response retrieval. More specifically, we break down the study into the following two concrete research questions:
\begin{itemize}
	\item \textbf{RQ1}: Since these tasks are all text based, does relevance modeling in different IR tasks really show differences in terms of natural language understanding?
	\item \textbf{RQ2}: If there do exist differences, how can we leverage these findings to enhance the relevance modeling on each IR task?
\end{itemize}

For the first question, we propose to leverage the probing-based method, which has been widely adopted in understanding the language modeling~\cite{conneau2018you,liu2019linguistic}, to analyze the potential differences in relevance modeling in the three IR tasks. Towards this goal, there are two basic requirements for the design of our empirical experiments: 1) It is better to have a unified IR model which can perform well on all these IR tasks, so that we can form a fair comparison basis. 2) The model should be able to integrate a variety of probing tasks, so that we can compare the modeling focuses easily. To meet these requirements, we take the recently proposed Bert model~\cite{devlin2018bert}, which have obtained reasonably good performances on these three retrieval tasks~\cite{dai2019deeper, nogueira2019passage, chen2019sequential}, as the unified IR model for study. We then utilize 16 probing tasks related to language modeling to compare the differences of relevance modeling in the three IR tasks from the language understanding perspective. 
For the second question, we utilize the intervention method to study how to enhance the relevance modeling in different IR tasks based on the previous findings. The basic idea is to interfere an existing relevance model with each probe task as an intervention factor to see how the performance varied on each retrieval task. 


Through the above experiments, our analysis reveals the following interesting results:
\begin{itemize}
	\item For RQ1: The answer is YES. The three IR tasks show different modeling focuses on relevance from the natural language understanding view.  Specifically, the document retrieval focuses more on semantic tasks, the answer retrieval pay attention to both syntactic and semantic tasks, while the response retrieval has little preference to most of the linguistic tasks. Beyond these differences, The understanding of the \textit{Synonym} seems universally useful for all the three retrieval tasks.

	\item Furthermore, we also find that there are different language understanding requirements for the two inputs in relevance models. A by-product is that we can thus analyze the \textit{inherent heterogeneity} of the IR task by comparing its modeling focuses on the two inputs. Through our analysis, it is interesting to find that the answer retrieval is the most heterogeneous one rather than the document retrieval which is often considered heterogeneous based on its surface form \cite{guo2019deep}.

	\item For RQ2: We demonstrate that we are able to improve the relevance modeling based on the above findings by the parameter intervention method. 
\end{itemize}

The rest of the paper is organized as follows: In Section 2, we describe the representative retrieval tasks in IR. We then present the probing analysis and intervention analysis in Section 3 and Section 4, respectively. The Section 5 discuss the related work while conclusions are made in Section 6. 

\section{Retrieval Tasks in Information Retrieval}
In this section, we introduce the IR tasks used in this work for the relevance modeling analysis.
Given a user's information need $S$ (e.g., query, utterance, or question), a retrieval task aims to find relevant information $T=\{t_1, t_2, ..., t_k\}$ (e.g., Web pages, response, and answers) from an archived information resources $\mathcal{T}$. Many applications can be formulated as an IR task, such as document retrieval, image retrieval, and so on. In this work, we focused on text-based retrieval tasks, and take three representative retrieval tasks for the relevance modeling analysis, namely document retrieval, answer retrieval, and response retrieval.


\subsection{Document Retrieval}
Document retrieval is a classical task in IR~\cite{voorhees2006overview}, which has been widely used in modern Web search engines, such as Google, Bing, Yandex, and so on. 
In this task, users typically specify their information needs via a query $Q$ to an information system to obtain the relevant documents $D$. The retrieved documents are returned as a ranking list through a ranking model according to their relevance degree to the input query. 
A major characteristic of document retrieval is the length heterogeneity between queries and documents. The user queries are often very short with unclear intents, consisting of only several key words in most cases. Existing works have shown that the average length of queries is about $2.35$ terms~\cite{10.1145/331403.331405}.
However, the documents are usually collected from the World Wide Web and have longer text lengths, ranging from multiple sentences to several paragraphs. 
This heterogeneity leads to the typical vocabulary mismatching problem, which has long been a challenge in the relevance modeling of document retrieval~\cite{li2014semantic}. 
To address this issue, a great amount of efforts has been devoted to design effective retrieval models to capture the semantic matching signals between the query and the document for document retrieval ~\cite{li2014semantic, guo2019deep}.

\subsection{Answer Retrieval}
Answer retrieval is widely used in question answering (QA) systems, such as StackOverflow~\footnote{https://stackoverflow.com/}, Quora~\footnote{https://quora.com/}, and Baidu Zhidao~\footnote{https://zhidao.baidu.com/}. 
The QA system directly retrieves the answer $A$ to the question $Q$ from existing answer repository $\mathcal{T}$. 
The core of the QA system is to compute relevance scores between questions and candidate answers, and subsequently ranking them according to the score. 
Compared with document retrieval, answer retrieval is more homogeneous and poses different challenges. Specifically, the questions are usually natural language, which are well-formed sentence(s) and have clearer intent description. While the answers are usually shorter text spans, e.g., sentences or passages, which have more concentrated topics.
However, answer retrieval is still a challenge problem since an answer should not only be topically related to but also correctly address the question.
Different retrieval models have been propose for the answer retrieval. Earlier statistical approaches focused on complex feature engineering, e.g., lexical and syntactic features~\cite{yih2013question}. In recent years, end-to-end neural models have been applied for relevance modeling in answer retrieval and achieved state-of-the-art performances~\cite{guo2019deep}.



\subsection{Response Retrieval}
Response retrieval is a core task in automatic conversation systems, such as Apple Siri, Google Now, and Microsoft XiaoIce. The conversation system relies on response retrieval to select a proper response $R$ from a dialog repository $\mathcal{T}$ with respect to an input utterance $U$. In multi-turn response retrieval, there is a context $C$ accomplished with each utterance $U$, where the context contains the conversation histories before the utterance.
Different from document retrieval and answer retrieval, the input utterance and candidate responses are often short sentence, which are homogeneous in the form. 
The relevance in response retrieval often refers to certain semantic correspondence (or coherent structure) which is broad in definition, e.g., given an input utterance "OMG I got myopia at such an 'old' age", the response could range from general (e.g., "Really?") to specific (e.g., "Yeah. Wish a pair of glasses as a gift") \cite{yan2016learning}. 
Therefore, it is often critical to model the coherence and avoid general trivial responses in response retrieval. In recently years, researchers have proposed a variety of approaches for response retrieval tasks~\cite{chen2017a}, where the neural network based methods have achieved state-of-the-art performance~\cite{chen2019sequential}.


\section{Probing Analysis}
\label{sec:task_description}

In this section, we aim to address the first research question, that is, \textit{whether the relevance modeling in different IR tasks really shows differences in terms of natural language understanding}. For this purpose, we propose to leverage the probing-based method to analyze the potential differences in the relevance modeling of the above three IR tasks. In the following, we will give the detailed description of the analysis process, including the probing method, the probing tasks, and the experimental results.

\subsection{The Probing Method}
The core idea of the probing analysis is to learn a unified representative retrieval model over the three IR tasks, and probe the learned model to compare the focuses between different relevance modeling tasks. 
Specifically, we take the recently proposed Bert model as the unified retrieval model since it has obtained reasonably good performances on all the retrieval tasks~\cite{dai2019deeper, nogueira2019passage, whang2019domain}. Moreover, the Bert model is a stack of multiple Transformer layers~\cite{devlin2018bert}, which can easily integrate different probing tasks on each Transformer layer. In this way, we could investigate the nuanced requirements of relevance modeling, and form a fair comparison between different IR tasks.

To learn the retrieval model for each IR task, we finetune the original Bert model to achieve good performances on each retrieval dataset respectively. 
We then probe the original Bert and the finetuned Bert with a set of natural language understanding tasks~\cite{belinkov2019analysis, liu2019linguistic}. 
Specifically, for the model to be probed, either the original or the finetuned Bert, we take an additional multi-layer perceptron (MLP) as the prediction layer over the target layers to be probed. We then train and evaluate the probing task over the model to assess its ability in capturing the corresponding linguistic properties.
It is worth to note that the Bert layers are fixed during the probing, since we aim to investigate what have been encoded in these layers.

Finally, we analyze the \textit{performance gap} of each probing task between the original and finetuned Bert over each IR task. Note that it is improper to directly compare the absolute performance of the finetuned Bert models on different IR tasks since the training corpus varies a lot. On the contrary, by taking the original Bert as a baseline, the relative performance gap of the finetuned Bert over the baseline on a probing task could reflect the importance of the specific linguistic property for the corresponding retrieval task.

\subsection{Probing Tasks}
We utilize a suite of 16 diverse probing tasks related to natural language understanding to investigate the focuses of the retrieval model, including lexical tasks, syntactic tasks, and semantic tasks. Here, most of the probing tasks have been utilized to study the linguistic properties of neural language models in different NLP tasks~\cite{belinkov2019analysis, liu2019linguistic}, e.g., language model~\cite{conneau2018you}, sentence embedding~\cite{perone2018evaluation}, natural language inference~\cite{richardson2019probing}. In this work, we take them to study the preferences of the relevance modeling in each retrieval task. 
In addition, we also introduce four probing tasks, which are closely related to the semantic matching between natural sentences, i.e., synonym identification, polysemy identification, keyword extraction, and topic classification. 
In the following, we will describe each probing task in detail, and the statistics of the datasets and the settings are listed in the Appendix \ref{appendix:probing_datasets}.


\subsubsection{Lexical Tasks}
The lexical tasks focus on the lexical meaning and positions of a term in sentences, paragraphs, or documents. It lies at the low level of the natural language understanding~\cite{pustejovsky2012semantics}. Here, we take three typical lexical tasks for the probing.

The \textbf{Text Chunking (Chunk)} task, also referred to as shallow parsing, aims to divide a complicated text into smaller parts. This task assesses whether the relevance modeling captures the notions of the spans and boundaries. We use the CoNLL 2000 dataset~\cite{sang2000introduction} for experiments.

The \textbf{Part-of-Speech Tagging (POS)} task is the process of marking up of words in a sentence as nouns, verbs, adjectives, adverbs etc. This task tests whether the relevance modeling captures the POS knowledge. Here, we take UD-EWT dataset ~\cite{silveira2014gold} for experiments.
 
The \textbf{Named Entity Recogntion (NER)} is the task of identifying entities and their category from a given text. This task assesses whether the relevance modeling pay attention to the entity information. We use the CoNLL 2003 dataset~\cite{sang2003introduction} for experiments.

\subsubsection{Syntactic Tasks}
The syntactic task is on the linguistic discipline dealing with the relationships between words in a sentence (i.e. clauses), the correct creation of the sentence structure and the word order.

The \textbf{Grammatical Error Detection (GED)} task is to detect grammatical errors in sentence. It is to assess whether the grammatical information is required for the relevance modeling. We use the First Certificate in English dataset~\cite{schneider2018comprehensive} for experiments.

The \textbf{Syntactic Dependence} task is to examine whether the syntactic relationships between words are crucial to model the relevance. We follow the work~\cite{liu2019linguistic} to take \textbf{arc prediction} and \textbf{arc classification} for experiments. Specifically, the \textit{syntactic arc dependency prediction} (\textbf{SynArcPred}) is a binary classification task, which aims to identify whether a relation exits between two tokens. The \textit{syntactic arc dependency classification} (\textbf{SynArcCls}) is a multi-class classification task, which assumes the input tokens is linked with each other and identifies which relationship it is. We use the UD-EWT datasets~\cite{silveira2014gold} for experiments.

The \textbf{Word Scramble} is a binary classification task which assesses whether the word order and structure affects the meaning of a sentence. We use it to test whether the relevance modeling cares about the work orders of the sentences/documents.
We use the PAWS-wiki dataset~\cite{zhang2019paws} for experiments.

\subsubsection{Semantic Tasks}
The semantic tasks deal with the semantic meaning of the words and sentences, the ways that words and sentences refer to each other. It lies at the high level of text understanding.

The \textbf{Preposition Supersense Disambiguation} is to examine whether semantic contribution of preposition is important factors to model the relevance. We follow previous work~\cite{liu2019linguistic} to take two sub-tasks for experiments, namely \textbf{PS-fxn} and \textbf{PS-role}. Specifically, the \textbf{PS-fxn} concerns the function of the preposition, while the \textbf{PS-role} determines the role of the preposition. We use the STREUSLE 4.0 corpus~\cite{schneider2018comprehensive} for experiments.

The \textbf{coreference arc prediction} (\textbf{CorefArcPred}) is to assess whether two mentions share the same coreference cluster. We use it to test whether the relevance modeling captures the coreference relationship between pronouns and entities.
We use the CoNLL dataset~\cite{pradhan2012conll} for experiments.

The \textbf{Semantic Dependence} task is to assess whether the semantic relationships between words are important for the relevance modeling. We follow the work~\cite{liu2019linguistic} to take \textit{arc prediction} and \textit{arc classification} for experiments. Specifically, the \textit{semantic arc dependency prediction} (\textbf{SemArcPred}) aims to identify whether a semantic relation exits between two tokens. The \textit{semantic arc dependency classification} (\textbf{SemArcCls}) assumes the input tokens is linked with each other and identifies which semantic relationship it is. We use the SemEval 2015 dataset~\cite{oepen2014semeval} for experiments.

The \textbf{Synonym} and the \textbf{Polysemy} task deal with the semantic meaning of a word pair from two sentences. The synonym focus on identifying whether two different words from similar context share the same meaning, while the polysemy is to distinguish the meaning of the same word from two sentences. We use them to test whether the relevance modeling captures the semantic meaning between word pairs.
For these two tasks, we crawled 10k sentences from an online Website for experiments. We will release these datasets after the paper is accepted.
The \textbf{Keyword} extraction task is to identify the prominent words that best describe the subject of a document. This task is to test whether the relevance modeling focuses on the keywords to interact the input pairs. Here, we take the Inspec~\cite{hulth2003improved} dataset for experiments.

The \textbf{Topic Classification} is to classify a document into a pre-defined topic. We use it to test whether the relevance modeling pay attention to the topics of the text inputs. Here, we use the Yahoo! Answers dataset~\cite{zhang2015character} for experiments since the topic categories are more suitable for the information retrieval applications.

\begin{table}[] 
\begin{tabular}{c c c c c c}
\toprule
Tasks & \#$S$ & \#$T$ & AvgLen($S$) & AvgLen($T$) & \#$\mathit{Vocab}$ \\
\hline
    Robust04  & 250 & 0.5M & 2.6 & 465 & 27194 \\
    MsMarco  & 100K & 1M & 6.4 & 56.3 & 27636 \\
    Ubuntu  & 0.59M & 0.66M & 10.3 & 22.2 &24026 \\
    \hline
\bottomrule
\end{tabular}
\caption{Dataset statistics of each retrieval tasks, $S$ represents the left input of each retrieval task. $T$ represents the right input of each retrieval task.}
\label{table:retrieval_datasets}
\end{table}

\subsection{Experimental setting}
For experiments, we first introduce the settings for the retrieval tasks, including models, datasets, and configurations. Then, we describe the settings of the probing tasks.

\begin{figure*}[!th]
\centering
\includegraphics[scale=0.65]{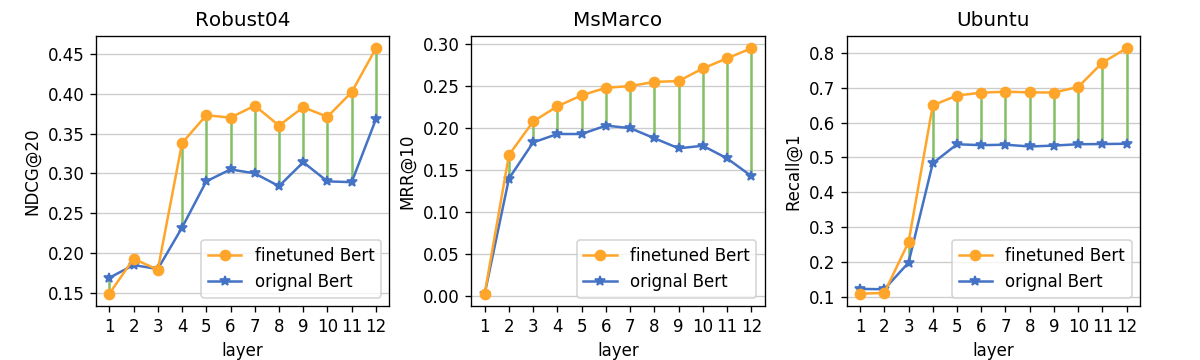}
\caption{Layer-wise performances of the original Bert and the finetuned Bert in different retrieval tasks.}
\label{fig:retrieval_task_performance}
\end{figure*}

\subsubsection{Retrieval model}
Here, we take the off-the-shell $BERT_{base}$ (~\citeauthor{devlin2018bert},~\citeyear{devlin2018bert}) model, which has been proved to be effective in many retrieval tasks~\cite{dai2019deeper,nogueira2019passage,whang2019domain}, as the retrieval model in all the three retrieval tasks. 
Specifically, the model takes the concatenation of a text pairs as input with a special token ``[SEP]'' separating the two segments. To further separate the left input from the right input, we follow the work~\cite{devlin2018bert} to add two additional tokens ``[S]'' and ``[T]'' into the two segments. 
All tokens are mapped into an embedding, with an additional position embedding concatenated in it. Then, the tokens go through several \textit{transformer} layers to fully interact with each other. Finally, the output embedding of the first token is used as the interaction of the input text pairs, and fed into a multi-layer perceptron (MLP) to obtain the final relevance score. 
For fair comparison, we directly take the Bert-base model\footnote{https://github.com/google-research/bert} (uncased, 12-layer, 768-hidden, 12-heads, 110M parameters) as the implementation as the retrieval model. For model learning, we leverage the pre-trained language model released in the original Bert as the initialization and finetune it on the corresponding datasets. The MLP layer is learned from scratch as the previous works~\cite{devlin2018bert}. For more detailed settings of all the retrieval models, please refer to Appendix \ref{appendix:retrieval_task_setting}.

\subsubsection{Retrieval Datasets}
To learn the retrieval model, we take three representative benchmark dataset, i.e., Robust04~\cite{voorhees2006overview}, MsMarco~\cite{bajaj2016ms}, and Ubuntu~\cite{lowe2015ubuntu}, for the relevance modeling in document retrieval, answer retrieval, and response retrieval, respectively. The statistics of these datasets are shown in Table~\ref{table:retrieval_datasets}. 
As we can see, these datasets show very different patterns in terms of the average length of the text pairs in different tasks. Document retrieval is the most heterogeneous as the average length of query and document is $2.6$ and $465$, respectively. While answer retrieval has reduced heterogeneity compared with document retrieval. The response retrieval is relatively homogeneous as the average length of the utterance and response are very close to each other. 
 For all these datasets, we simply padded each short text pairs with [PAD] and truncated long text pairs into $512$ tokens. 
 
 For task evaluation, we take the NDCG@20 for document retrieval, MRR@10 in answer retrieval, and recall@1 in response retrieval, as is done in previous works~\cite{dai2019deeper,nogueira2019passage,whang2019domain}.




\subsection{Results}
In this section, we show the probe experiments as well as the results by answering the following research questions.

\subsubsection{\textbf{How does the unified retrieval model perform on each retrieval task?}}\label{retrieval_performance_section}
We take the same original Bert model as the starting point, and finetune it on each IR task to learn task specific requirements for relevance modeling. 
In the following sections, we will use BERT$_\textit{base}$ to denote the original Bert model, and BERT$_\textit{doc}$, BERT$_\textit{ans}$, and BERT$_\textit{rsp}$ to denote the finetuned Bert on document retrieval, answer retrieval, and response retrieval, respectively.
Here, we show the performances of the BERT$_\textit{base}$ as well as the finetuned models on the IR datasets with respect to each layer in Figure~\ref{fig:retrieval_task_performance}. 
The results are summarized as follows.

Firstly, we can see that the BERT$_\textit{base}$, which learned over a large amount of unstructured texts in an unsupervised way, has already achieved good performances on all the three retrieval tasks (The existing state-of-the-art performances on each datasets are listed in appendix \ref{appendix:retrieval_results}). 
It indicates that the linguistic information encoded in Bert~\cite{liu2019linguistic} is useful for relevance modeling. In addition, it is worth to note that the best performance of the BERT$_\textit{base}$ is not always achieved at the last layer, e.g., the answer retrieval on MsMarco gets the best result on the sixth layer. The results suggest that the probing should better be conducted over all the layers, not just the last layer, to select the best-performing layer to study.

Secondly, we can see that the finetuned Bert can significantly improve the performances on all the retrieval tasks. Specifically, the relative improvement of the fintuned Bert (i.e., BERT$_\textit{doc}$, BERT$_\textit{ans}$, and BERT$_\textit{rsp}$) against the BERT$_\textit{base}$ over the best layer is about $23.3\%$, $45.3\%$, and $51.6\%$, respectively.
These improvements indicate that the finetuned Bert models are able to learn task specific properties for the relevance modeling on each IR task.

Finally, we can observe that the finetuned Bert achieved larger improvements on the higher layers than the lower layers on all the three tasks, and the last layer always performs the best. 
This is consistent with the findings of existing work~\cite{liu2019linguistic} that the higher layers of the finetuned Bert tend to learn the task specific features, while the lower layers learn the basic linguistic features.

\begin{table*}[t] 
\caption{Overall performances of each probing tasks on different retrieval tasks. Significant improvement or degradation with respect to Bert$_\textit{base}$ is indicated (+/-) ($\text{p-value} \le 0.05 $ with Bonferroni correction)).}
\begin{tabular}{cl | ccc | ccc | ccc}
\toprule
\multicolumn{2}{c|}{Probing Tasks} & \multicolumn{3}{c|}{Document Retrieval} & \multicolumn{3}{c|}{Answer Retrieval} & \multicolumn{3}{c}{Response Retrieval} \\
\hline
& & base & Bert$_\mathit{doc}$ & $\Delta$ & base & Bert$_\mathit{ans}$ & $\Delta$ & base & Bert$_\mathit{rsp}$ & $\Delta$ \\
\hline
 \multirow{3}{2.4cm}{\centering Lexical Tasks} &
    Chunk &92.6 &92.47 & -0.14\% &92.9 & 92.53& -0.40\% & 92.47 & 92.49 &  +0.02\%\\
   & POS  & 95.89&95.72 &  -0.18\% &95.45 &95.48 &  +0.03\% & 95.7 & 95.55 &  -0.16\%\\
   & NER  &83.51 & 83.16&  -0.42\%$^{-}$ &80.1 &80.95 &  +1.06\%$^{+}$ &82.16 & 80.72&  -1.52\%$^{-}$\\
    \hline
\multirow{4}{2.4cm}{\centering Syntactic Tasks} &
    GED & 41.83 & 40.56 &  -3.04\%$^{-}$ &41.44 & 41.8&  +0.87\% & 41.23&39.72 &  -3.66\%$^{-}$  \\
  &  SynArcPred  &87.29 &87.21 &  -0.09\% &86.44 & 86.75&  +0.36\% &86.44 &85.95 &  -0.57\%$^{-}$\\
  &  SynArcCls  & 93.66&93.59 &  -0.07\% &93.37 & 93.43&  +0.06\% &93.32 &92.87 &  -0.25\%\\
  &  Word Scramble & 62.04&61.87 &  -0.27\%$^{-}$ &62.17 &62.87 &  +1.13\%$^{+}$ & 59.91&60.19 &  +0.47\%$^{+}$\\
    \hline
\multirow{9}{2.4cm}{\centering Semantic Tasks} &
    PS-fxn  &89.21 &89.89 &  +0.76\%$^{+}$ &87.92 & 88.95&  +1.17\%$^{+}$ &89.95 &86.89 &  -3.4\%$^{-}$\\
  &  PS-role  & 78.2 &79.55 &  +1.73\%$^{+}$ & 77.63 & 79.18& +2.00\%$^{+}$ &79.22 & 80.14&  +1.16\%$^{+}$\\
  &  CorefArcPred  &78.22 & 78.46&  +0.31\%$^{+}$ & 77.5& 76.93&  -0.74\%$^{-}$ & 79.53& 78.3&  -1.6\%$^{-}$\\
  &  SemArcPred  &87.34&86.96 &  -0.44\%$^{-}$ &87.69 & 88.01&  +0.06\% &87.23 &87.09 &  -0.16\%\\
  &  SemArcCls  &92.47 &92.45 &  -0.02\% & 92.67& 92.43 &  -0.43\%$^{-}$&92.98 &92.33 &  -0.7\%$^{-}$\\
  &  Polysemy & 64.1 &67.1 &  +4.68\%$^{+}$ &64.1 &69.1 &  +7.8\%$^{+}$ &64.1 &58.9 &  -11.17\%$^{-}$\\
  &  Synonym &66.32 &78.49 &  +18.35\%$^{+}$ &66.33 &75.86 &  +14.37\%$^{+}$ &66.31 &80.68 &  +21.67\%$^{+}$\\
  &  Keyword &48.66 &48.98 &  +0.66\%$^{+}$ &48.84 &48.72 &  -0.25\%&46.66 &45.95 &  -1.52\%$^{-}$\\
  &  Topic &66.93 &67.82 &  +1.33\%$^{+}$ &67.7 &69.16 &  +2.16\%$^{+}$ &67.34 &66.11 &  -1.83\%$^{-}$ \\
    \hline
\bottomrule
\end{tabular}
\label{table:prob_result}
\end{table*}

\subsubsection{\textbf{Do different IR tasks show different modeling focuses in terms of natural language understanding?}}
\label{sec:prob_result}
Here, we study the differences between IR tasks through quantitative analysis based on the performance gap of the probing tasks.
As found in previous section, the best performance of a probing task could be achieved by any layer of the Bert model. For fair comparison, we take the best layer from the BERT$_\textit{base}$ model and the finetuned Bert models (i.e., BERT$_\textit{doc}$, BERT$_\textit{ans}$, and BERT$_\textit{rsp}$) for the following study. 
The results are summarized in the Table~\ref{table:prob_result}. 

We first look at each IR task and find the following performance patterns. 
\begin{itemize}
    \item[1)] For document retrieval, there is a clear pattern that the relevance modeling focuses more on the semantic tasks than the lexical and syntactic tasks. The performance gap on most semantic tasks between  BERT$_\mathit{doc}$ and BERT$_\textit{base}$ is positive and significant. Among them, The top-2 improved tasks are Synonym and Polysemy, showing that relevance modeling in document retrieval requires better understanding of the semantic meaning of a word pair. This is somehow consistent with the previous findings~\cite{DBLP:conf/ecir/YiA09} that topic models (e.g. PLSI~\cite{DBLP:conf/sigir/Hofmann99} and LDA~\cite{DBLP:conf/sigir/WeiC06}), which capture the synonym and polysemy well, can be applied to improve the document retrieval models. 
    \item[2)] For answer retrieval, most probing tasks (i.e., 11 out of 16) have been improved by BERT$_\mathit{ans}$, among which eight improvements are significant. It indicates that the relevance modeling in answer retrieval is more difficult, which requires more comprehensive language understanding as compared with the other two. Specifically, BERT$_\mathit{ans}$ improves all the syntactic-levels tasks, showing that the syntactic features, like word order and structure in a sentence, are important for relevance modeling in answer retrieval. 
    \item[3)] For response retrieval, it is surprising to see that the performances of most probing tasks (i.e., 12 out of 16) have been decreased by Bert$_\mathit{rsp}$, among which ten drops are significant. It suggests that most linguistic properties encoded by the original Bert has already been sufficient for the relevance modeling in response retrieval. Meanwhile, we can find that Bert$_\mathit{rsp}$ improves Synonym while decreases Polysemy significantly, as two extremes. The results demonstrate that response retrieval need to better understand similar words in different contexts than to distinguish the same words in different context. 
\end{itemize}

We then look at each probing task and obtain the following observations across different IR tasks.
\begin{itemize}
    \item[1)] The \textit{CorefArcPred} and \textit{Keyword} tasks have only been significantly improved by BERT$_\mathit{doc}$ among the three finetuned models but decreased by the rest. Meanwhile, the \textit{NER} and \textit{GED} tasks have only been significantly improved by BERT$_\mathit{ans}$ but drop on the other two. The results indicate that relevance modeling in document retrieval pays more attention to similar keywords while the relevance modeling in answer retrieval pays more attention to identifying targeted entities in questions and answers.
    \item[2)] The \textit{Word Scramble} task obtains significant improvement by both BERT$_\mathit{ans}$ and BERT$_\mathit{rsp}$ but drops by BERT$_\mathit{doc}$. It suggests that the relevance modeling in both answer retrieval and response retrieval cares more about the word order and sentence structure than that in document retrieval. This also explains why keyword-based methods could work very well for ad-hoc retrieval (i.e., document retrieval). Moreover, the \textit{Polysemy} and \textit{Topic} tasks obtain significant improvement by BERT$_\mathit{doc}$ and BERT$_\mathit{ans}$, but drop significantly  BERT$_\mathit{rsp}$.  In fact, the \textit{Polysemy} is also connected with topic identification since it aims to identify polysemic words under different topics. This indicates that the relevance modeling in both document retrieval and answer retrieval pays more attention to topic understanding than that in response retrieval.
    \item[3)] Despite the above differences, there are some common patterns across the three tasks. We can see that both the \textit{Synonym} and \textit{PS-role} tasks have been improved significantly by all the three finetuned Bert models. Moreover, the improvement of the \textit{Synonym} task is always the largest on all the three retrieval tasks. These results demonstrate that it is of great importance to capture the synonyms in all the relevance modeling tasks. 
\end{itemize}

Based on all the above observations, we can conclude that the relevance modeling in the three representative retrieval tasks shows quite different modeling focuses in terms of natural language understanding.

\begin{figure*}[!tb]
\centering
\includegraphics[scale=0.58]{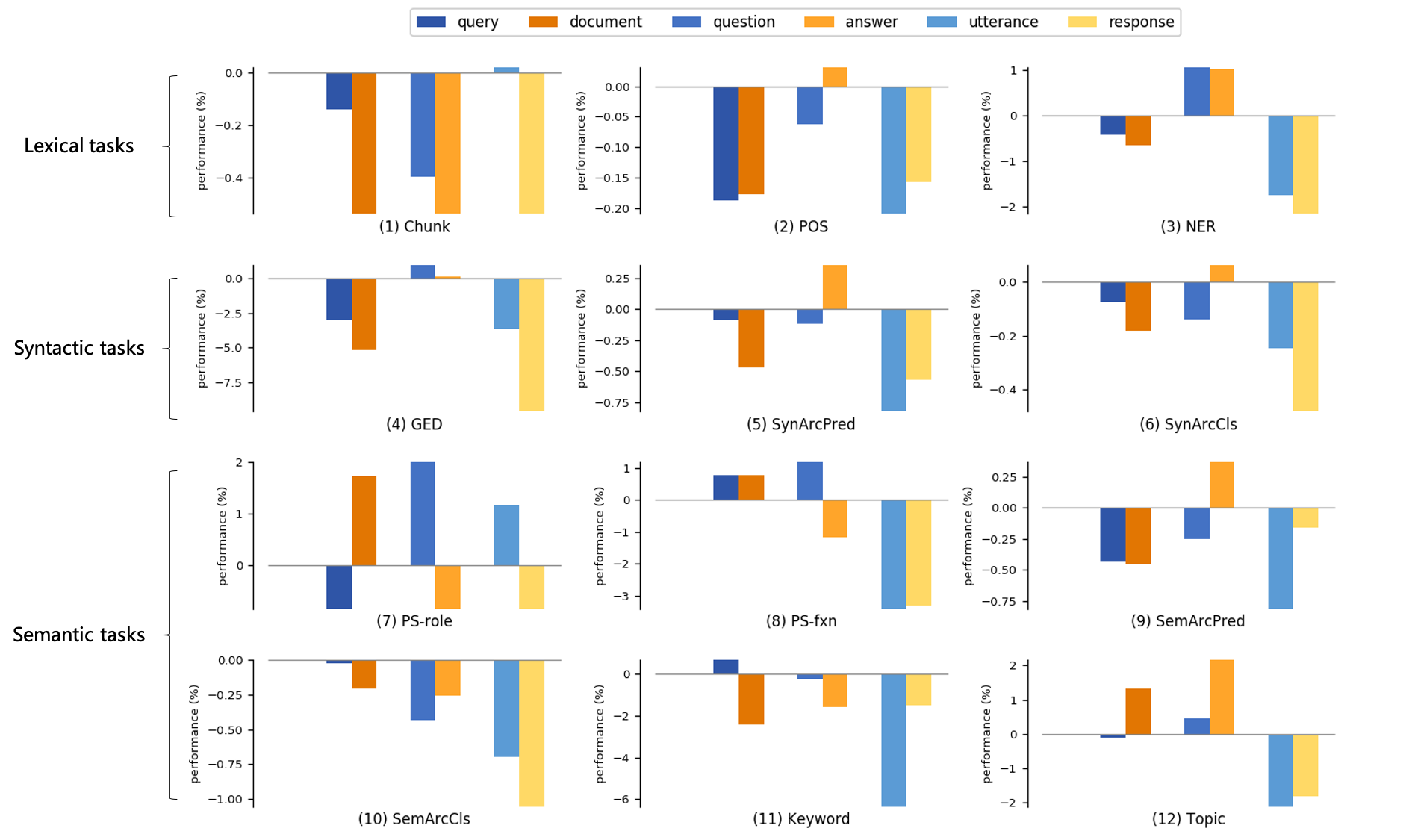}
\caption{Probing task performance comparison between the left and right input on three retrieval tasks. Each bar denotes the improvement/decrease in the performance over the corresponding baseline. The \textit{query}/\textit{question}/\textit{utterance} and \textit{document}/\textit{answer}/\textit{response} are the left and right input of document retrieval, answer retrieval, and response retrieval, respectively.}
\label{fig:left_right_bar}
\end{figure*}

\subsubsection{\textbf{Do relevance modeling treat their inputs differently in terms of natural language understanding?}}
Since relevance models typically take a pair of texts as inputs, 
we further study the performance gap of each probing task on the left and right input, respectively.
Here we directly mask the tokens in the right input when testing the left input, and vice versa. 
In this study, we only keep the probing tasks whose input is a single sentence, and ignore the tasks that require a pair of inputs (i.e., The \textit{Word Scramble}, \textit{CorefArcPred}, \textit{Polysemy}, and \textit{Synonym}).
Similar as the previous section, we take the best layer in BERT as the representative performance for each comparison. The results are depicted in Figure~\ref{fig:left_right_bar}, with the blue and orange bar represent the performance gap between the BERT$_\textit{base}$ model and the finetuned Bert models on the left and right input, respectively. In the following, we use the term ``similar trend'' to denote the case where the performance gap are both positive or both negative on the left and right inputs, and use the term ``reverse trend'' if the gap directions are reverse to each other on the two inputs.
Specifically, we have the following observations: 
\begin{itemize}
    \item[1)]  In document retrieval, the left (query) side and the right (document) side show similar trends on both lexical and syntactic probing tasks, although the gap sizes are different in most cases. Meanwhile, they also show different modeling focuses on most semantic tasks. Specifically, the query side cares about the coarse-level functions of the prepositions (i.e., \textit{PS-fxn}) while the document side pays attention to both the coarse-level functions and the fine-grained roles (i.e., \textit{PS-fxn} and \textit{PS-role}) in terms of the prepositions. The query side improves the performance  on \textit{Keyword} but drops on \textit{Topic}, while the document side does the opposite. This is reasonable since queries are usually short keywords while documents are typically long articles.
    \item[2)] In answer retrieval, we can see that the left (question) side and the right (answer) side show very different preferences on most probing tasks. The question side improves $5$ out of $12$ probing tasks, while the answer side improves $7$ out of $12$ probing tasks. More importantly, they show the reverse trend on half of the probing tasks (i.e., \textit{POS}, \textit{SynArcPred}, \textit{SynArcCls}, \textit{PS-role}, \textit{PS-fxn} and \textit{SemArcPred}). Moreover, we can find that the question side pays more attention to semantic tasks, while the answer side cares more about the lexical and syntactic tasks. The results also indicate that understanding prepositions properly (i.e., \textit{PS-role} and the \textit{PS-fxn}) could be of great importance in understanding the question.
    \item[3)] In response retrieval, the left (utterance) side and the right (response) side show similar trends on most of the probing tasks (i.e. $10$ out of $12$), with  \textit{Chunk} and \textit{PS-role} as the exceptions. Among those similar trends, the gap sizes are quite different on the two sides. For example, the utterance side drops more on \textit{POS}, \textit{SynArcPred}, \textit{Keyword} and \textit{SemArcPred},  while the response side drops more on \textit{Chunk}, \textit{GED}, and \textit{SynArcCls}.
\end{itemize}

Based on the above results, we can further analyze the \textit{inherent heterogeneity} of the three retrieval tasks by comparing the linguistic focuses between their left and right inputs.
Here we take the reverse trend on the two inputs as a key signal for the inherent heterogeneity, which indicates significantly different modeling focuses on a probing task. As a result, we can find that the answer retrieval (i.e., $6$ reverse trends) is the most heterogeneous inherently from the linguistic view, followed by the document retrieval (i.e., $3$ reverse trends) and the response retrieval (i.e., $2$ reverse trends). This is an interesting result since the previous works \cite{guo2019deep, DBLP:conf/kdd/YinHTDZOCKDNLC16} often deem the document retrieval the most heterogeneous task due to the significant surface length and linguistic form differences between its inputs (i.e., the query and the document). Now from the natural language understanding view, we show that answer retrieval is more heterogeneous since it requires quite different understanding abilities on its two inputs.

\section{Intervention Analysis}\label{intervention_section}
In this section, we further study whether the previous findings on the differences of relevance modeling could actually give us some guidelines on model improvement. 
Inspired by the causal analysis~\cite{eberhardt2007interventions}, we take the intervention method to study whether some language understanding task could really help to improve the relevance modeling. 
The core of the intervention method is to take the probing task as the causal factor to interfere the retrieval model, and analyze the change of performances before and after the intervention. 
Specifically, we first learn the relevance model on each retrieval dataset to obtain the basic results for comparison. Then, we take either features or labels of each intervention factor on the same retrieval dataset to interfere the learning process of the relevance model with other factors hold fixed, and evaluate the performance of intervened models.
In the following, we will introduce intervention settings and experimental results in detail.

\subsection{Intervention Settings}
Here, we choose four representative probing tasks as intervention factors, i.e., \textit{Keyword}, \textit{NER}, \textit{Synonym}, and \textit{SemArcCls}, which is based on the following observations: 1) The \textit{Synonym} has shown to be consistently improved on all the three retrieval tasks. 2) The \textit{SemArcCls} has shown to be consistently deceased on all the three retrieval tasks. 3) The \textit{Keyword} and the \textit{NER} tasks have obtained distinct improvement on the document retrieval and answer retrieval, respectively. 
It is worth to note that the intervention process require the retrieval dataset to contain the label of each intervention factor, which would take enormous workloads to obtain the groundtruth label.
Recently, the weak labeling method has attracted considerable attention and shown to be beneficial in many NLP tasks~\cite{hoffmann2011knowledge}. 
Therefore, we take the finetuned Bert large~\footnote{https://github.com/google-research/bert}, which has been proved to be effective in all four intervention factors, to generate weak labels for each instance in all three retrieval datasets (i.e., Robust04, MsMarco, and Ubuntu). Then, the label of each intervention factor is used to interfere the learning process of the retrieval model. 
The details of each intervention method are described as follows:
\begin{itemize}
	\item \textit{Feature Intervention}: For feature intervention, we take the label of each instance as an additional input to the retrieval model. Specifically, we map the label of each factor (e.g., PER, ORG, LOC in the \textit{NER}) to embedding space and add the feature embedding to the BERT input embeddings. Thus, the final input embeddings of the retrieval model are the sum of the token embeddings, the segmentation embeddings, the position embeddings, and the feature embeddings.
Here, the embedding size of each feature is set to $768$ as is in original Bert model.
	\item \textit{Parameter Intervention}: For parameter intervention, we  firstly learn the retrieval model using the label of each intervention factor as the initial parameter, and then finetune the parameters of the model with each retrieval dataset. It is worth to note that we add an additional multi-layer perceptron layer on top of the relevance model to adapt it for each intervention factor. In the experiments, we learn each intervention factor with a small learning rate of $1e-5$, and finetune on the retrieval task with learning rate of $3e-5$.
	\item \textit{Objective Intervention}: For objective intervention, we jointly learn the intervention factor as well as the retrieval task. For this purpose, we add a task-specific layer on top of the Bert model for each intervention factor. For example, we add a CRF layer on top of the Bert for sequence labeling tasks (i.e., \textit{NER}), and add a linear layer on top of the Bert for classification tasks (i.e., \textit{Keyword}, \textit{SemArcCls}, and \textit{Synonym}).
	The loss function is a weighted sum of the ranking cross-entropy function and factor-specific loss function:  
$$ Loss = \lambda Loss_\mathit{ranking} + (1-\lambda)Loss_\mathit{factor},$$
where the $\lambda$ is learned in an end-to-end way. 
\end{itemize}

\begin{table}[!t] 
\caption{Results of different intervention methods based on the \textit{Keyword} task on different retrieval models. BERT$_\mathit{doc}$, BERT$_\mathit{ans}$, and BERT$_\mathit{rsp}$ is the finetuned Bert on document retrieval, answer retrieval, and response retrieval, respectively. Significant improvement or degradation with respect to Bert$_\textit{base}$ is indicated (+/-) ($\text{p-value} \le 0.05 $).}
\begin{tabular}{c ccc}
\toprule
\multirow{2}{1.6cm}{\centering intervention type} & BERT$_\mathit{doc}$ &BERT$_\mathit{ans}$ &BERT$_\mathit{rsp}$ \\
\cline{2-4}
& 0.459 & 0.367 & 0.817 \\
 \hline
 feature & 0.457 (-0.4\%) & 0.367 (-) & 0.810 (-0.1\%) \\
\cline{1-4}
parameter & 0.468 (+2\%$^{+}$)  & 0.355 (-11.7\%$^{-}$)  & 0.721 (-3.3\%$^{-}$) \\
 \cline{1-4}
objective & 0.402 (-12.4\%$^{-}$) & 0.341 (-8.7\%$^{-}$) & 0.746 (-7.1\%$^{-}$) \\
\bottomrule
\end{tabular}
\label{table:diff_intervention_results}
\end{table}

\subsection{Results}
 In this section, we show the intervention results of each intervention factors, including the comparison of different intervention methods and the analysis of different intervention factors.
 
\subsubsection{Intervention Methods Comparison}
Here, we compare each intervention method on all the three retrieval tasks based on the intervention factor of \textit{Keyword}.
The overall results are summarized in Table~\ref{table:diff_intervention_results}.
Firstly, we can see the that the feature intervention has very little effect on the performances of all the three retrieval tasks. This maybe that the embedding features of each token, which are built on the corresponding intervention factor, are not much effective for the retrieval modeling. 
Secondly, the objective intervention has significantly decreased the retrieval performances on all the three retrieval tasks with a large margin. The reason may be that the multi-task learning could possibly introduce inductive bias, which would lead to sub-optimal performances on individual tasks~\cite{aljundi2017expert}.  
Finally, the parameter intervention has gained significant improvements on the document retrieval task, and dropped with a large margin on the answer retrieval task and the response retrieval task. This is consistent with the previous findings on the probing analysis section, and verifies the importance of keyword recognization in the retrieval modeling of document retrieval.
All these results demonstrate that the parameter intervention is more effective than the other two intervention methods.

\begin{table}[!t] 
\caption{Results of different intervention factors using the \textit{parameter intervention} on each retrieval model. Significant improvement or degradation with respect to the finetuned Bert on the corresponding retrieval task is indicated (+/-) ($\text{p-value} \le 0.05 $).}
\begin{tabular}{c ccc}
\toprule
\multirow{2}{*}{Baseline} & BERT$_\textit{doc}$ & BERT$_\textit{ans}$ & BERT$_\textit{rsp}$ \\
\cline{2-4}
& 45.9 & 36.7 & 81.7\\
 \hline
   NER & 45.3 (-1.33\%$^{-}$) & 38.5 (+4.76\%$^{+}$) & 80.9 (-0.98\%$^{-}$) \\
 \cline{1-4}
Keyword &46.8 (+1.94\%$^{+}$) &35.5 (-3.35\%$^{-}$)  &72.1 (-11.75\%$^{-}$) \\
\cline{1-4}
SemArcCls & 39.1 (-21.78\%$^{-}$) & 26.7(-14.86\%$^{-}$)  & 63.9(-27.45\%$^{-}$) \\
\cline{1-4}
Synonym &46.3 (+0.83\%$^{+}$) &37.0 (+0.5\%) & 82.6 (+1.1\%$^{+}$) \\
\bottomrule
\end{tabular}
\label{table:intervent_factor_strong}
\end{table}

\subsubsection{Intervention Factors Analysis}
In this section, we further study whether and how different intervention factors could improve the relevance modeling through the parameter intervention. 
The intervention results are summarized in Table~\ref{table:intervent_factor_strong}. 
Firstly, we can see that the \textit{NER} and \textit{Keyword} have significantly improved the performances of the retrieval model in answer retrieval and document retrieval, respectively. For example, the \textit{NER} improved the BERT$_\mathit{ans}$ with a large margin to $4.76\%$. This verifies that it is of great importance to capture the entity information for relevance modeling in the answer retrieval.
Secondly, the \textit{SemArcCls} has unsurprisingly reduced the performances of the retrieval model on all the three retrieval tasks, which is also consistent with the findings in the probing analysis section. This demonstrates that it is not much useful to model the semantic dependencies between words for relevance modeling in these three retrieval tasks.
Finally, it can be observed that the \textit{Synonym} has consistently improved the performances of all relevance models. For example, the relative improvements of each retrieval model are $0.83\%$, $0.5\%$, and $1.1\%$ on document retrieval, answer retrieval, and response retreival, respectively. 
Moreover, it is worthy to note that the labels are automatically generated by an effective model of each intervention factor, which are often somewhat noisy and uncertain. Thus, it would be expected to further enhance the retrieval model if there exists ground-truth labels for each intervention factor.
All these results demonstrate that the factors revealed in the probing analysis could really be helpful to the relevance modeling for different retrieval task.

\section{Related Works}
In this section, we will introduce the works related to our study, including the relevance modeling and the probing analysis.
%

\subsection{The Relevance Modeling}
Relevance modeling is a core research problem in information retrieval. During the past decades, researchers have proposed a numerous number of relevance models for different retrieval tasks. 
In the document retrieval, different kinds of methods have been proposed to measure the relevance between a query and a document ~\cite{harman2019information}, including traditional heuristic methods~\cite{robertson1976relevance,Salton1975}, learning to rank methods~\cite{joachims2006training, burges2010ranknet}, and neural ranking methods~\cite{guo2019deep}. Firstly, the traditional heuristic methods, such as BM25, and TFIDF, build the heuristic function based on term frequencies, document length, and term importance.
 Then, the learning to rank models try to learn the ranking function based on machine learning methods on human designed features. Finally, the neural models, which attracted a great attentions in recent years, automatically learn the ranking function as well as the features based on neural networks~\cite{Onal2018neural}.
In answer retrieval, different methods are proposed to model the relevance relationship between the questions and the answers~\cite{murata1999question, mitra2017neural, guo2019deep}. In early days, traditional methods focused on the feature engineering, such as lexical, syntactic, and semantic features~\cite{murata1999question, yih2013question}. Recently, deep learning methods have significantly improved the answer ranking tasks, and become a mainstream method in this task~\cite{yang2016anmm,mitra2017neural,DBLP:conf/acl/LeeCT19}. 
In response retrieval, the relevance model is designed to evaluate the relevance degree between the utterances and the responses. Early methods was designed with handcrafted templates and heuristic rules. In recent years, neural models have became the mainstream along with the large scale human-human conversation data available~\cite{DBLP:conf/sigir/YangQQGZCHC18,chen2019sequential,whang2019domain}, and pushed forward the development of the conversation systems.

Though numerous relevance models have been introduced by considering requirements under different retrieval tasks, there has few works try to analyze the relevance modeling under different retrieval applications. To the best of our knowledge, this is the first work to study the relevance modeling in different retrieval tasks based on the empirical analysis.

\subsection{The Probing Analysis}
Recently, the probing tasks have been widely used to understand the powerful neural models, since the neural models often serve as a black-box in the downstream tasks. The core of the probing methods is to apply the linguistic tasks on the target model to investigate the properties based on the performance of these tasks.
A number of works have been proposed to study the linguistic properties of the learned representations over neural models \cite{belinkov2017neural, conneau2018you, peters2018dissecting}. For example, Belinkov et al.~\cite{belinkov2017neural} investigated the morphology information through several probing tasks like part-of-speech and semantic tagging on neural MT models. Conneau et al.~\cite{conneau2018you} constructed a set of probing tasks to study the linguistic properties of sentence embedding. 
There are also several works try to investigate how to design a good probe for the model understanding \cite{DBLP:journals/corr/abs-1809-10040, DBLP:conf/acl/WuCKL20}. For example, Zhang et al.~\cite{DBLP:journals/corr/abs-1809-10040} presented experiments to understand how the training sample size and memorization affect the performance of linguistic tasks.
Hewitt and Liang~\cite{DBLP:conf/naacl/HewittM19} investigated the selectivity of probes, and proposed control tasks to study the expressivity of probe tasks and methods. 

These studies inspired us to analyze the relevance modeling in different retrieval tasks. However, most of the existing works take the probing tasks directly to investigate the property of the pre-trained language model, we instead use them to compare the focuses of retrieval tasks under different applications.



\section{Conclusion}
In this paper, we present an empirical analysis of the relevance modeling in different retrieval tasks, including document retrieval, answer retrieval, and response retrieval. We propose to use the probing method to investigate the relevance modeling, and introduce 16 probing tasks for the relevance analysis. The results show some interesting findings about the focuses of different retrieval tasks.
To further study how to leverage these findings to improve the relevance modeling in each retrieval task, we introduce three intervention methods, i.e., feature intervention, parameter intervention, and objective intervention, to interfere existing retrieval models. The intervention results demonstrate that it is able to improve the retrieval models based on the findings on language understanding by carefully designed intervention methods.
The analysis of the relevance modeling is a foundation for designing effective relevance models in real world applications. We believe the way we study the problem (probing \& intervention) as well as our findings would be beneficial to the IR community.
For future work, we'd like to apply the findings of the probing analysis to improve existing retrieval models. Moreover,  we would also try to design new effective retrieval models based on the findings in this work.

\section{Acknowledgments}
This work was supported by the National Natural Science Foundation of China (NSFC) under Grants No. 61902381, 61722211, 61773362, and 61872338, and funded by Beijing Academy of Artificial Intelligence (BAAI) under Grants No. BAAI2019ZD0306 and BAAI2020ZJ0303, the Youth Innovation Promotion Association CAS under Grants No. 20144310, and 2016102, the National Key RD Program of China under Grants No. 2016QY02D0405, the Lenovo-CAS Joint Lab Youth Scientist Project, the K.C.Wong Education Foundation, and the Foundation and Frontier Research Key Program of Chongqing Science and Technology Commission (No. cstc2017jcyjBX0059).



%
\appendix
 \section{Retrieval Task Setting}
 \label{appendix:retrieval_task_setting}
To learn the relevance model for each retrieval task, here, we following existing works to apply the corresponding ranking loss on each datasets. Specifically, for Robust04 dataset, we utilize the pairwise ranking loss (i.e., hinge loss)~\cite{dai2019deeper} to train the retrieval model, i.e., given a triple $(s, t^+, t^-)$, where $t^+$ is ranked higher than $t^-$ with respect to a query $s$, the loss function is defined as:
$$\mathcal{L}(s, T^+, T^-; \theta) = \sum_{t^+ \in T^+, t^- \in T^-}\max(0, 1-f(s, t^+) + s(s, t^-)),$$
where $f(s, t)$ denotes the relevance score for pair $(s, t)$, and $\theta$ includes all learnable parameters in the ranking model.
For Msmarco and Ubuntu dataset, we take the cross entropy loss~\cite{nogueira2019passage, whang2019domain} to train the retrieval model,
$$\mathcal{L}({s, T^+, T^-, \theta})  = -\sum_{j \in T^+}\log(f(s, t_j)) - \sum_{j \in T^-}\log(f(s, t_j)),$$
where $T^+$ and $T^-$ denote the positive and negative answers/responses with respect to question/utterance $s$.
The optimization is relatively straightforward with standard backpropagation. We apply stochastic gradient decent method Adam with learning rate warmup over the first 10\% training steps, and linear decay of the learning rate for BERT layers, fixed learning rate of 0.001 for task-specific layer which always be a linear classification layer. We use a dropout probability of 0.1 on all layers. 
Since the length of text inputs in each retrieval tasks differ significantly wit each other, we thus tailor the input length for each dataset accordingly. Specially, for Robust04 dataset, the maximum sequence length is set to 512 where the length of left and right input is set to 30 and 480, respectively. For MSMARCO dataset, the maximum sequence length is set to 230, where the length of left and right input is set to 30 and 200, respectively. For Ubuntu dataset, the maximum sequence length is set to 300, where the length of left and right input is set to 256 and 44, respectively.

\section{References to State-of-the-Art Models on Three Retrieval Tasks}
\label{appendix:retrieval_results}
\begin{table}[!h] \label{References to State-of-the-Art Models on Retrieval Tasks}
\begin{tabular}{l l l}
\toprule
\multirow{1}{*}{Task} & \multirow{1}{*}{Previous state of the art} & \multirow{1}{*}{BERT} \\\hline
\multirow{1}{*}{Robust04} &43.1 (\citeauthor{zamani2018neural},\citeyear{zamani2018neural}) & 46.9 (\citeauthor{dai2019deeper},\citeyear{dai2019deeper}) \\
\multirow{1}{*}{MsMarco} & 27.1 (\citeauthor{dai2018convolutional}, \citeyear{dai2018convolutional})& 35.8 (\citeauthor{nogueira2019passage},\citeyear{nogueira2019passage}) \\
\multirow{1}{*}{Ubuntu} & 79.6 (\citeauthor{chen2019sequential}, \citeyear{chen2019sequential}) & 81.7 (\citeauthor{whang2019domain},\citeyear{whang2019domain})\\
\bottomrule
\end{tabular}
\caption{A Comparison of Performance of prior state of the art models and BERT.}
\label{table:retrieval_sota}
\end{table}

 \section{Probing Settings}
 \label{appendix:probing_datasets}
 The statistics of the dataset used in each probing tasks are listed in Table \ref{table:probe_datasets}.
For all the probing experiments, we follow existing works to use linear probe on all these tasks as it has been proved to have better selectivity~\cite{hewitt2019designing}. Specifically, we add a linear layer on top of each layer in the Bert model as the prediction layer of each probing task.
We follow ~\citeauthor{liu2019linguistic}~\citeyear{liu2019linguistic}'s work and take the \emph{contextual-repr-analysis}~\footnote{https://github.com/nelson-liu/contextual-repr-analysis} toolkit for the probing experiments. This toolkit is implemented under the AllenNLP(~\citeauthor{Gardner2017AllenNLP},~\citeyear{Gardner2017AllenNLP}) framework. The description of the probe datasets are listed in the Table~\ref{table:probe_datasets}. Note that we build two novel probing tasks, i.e., Synonym and Polysemy, to directly evaluate the semantic understanding of word pairs.
For performance evaluation, we take the F1 metric for Chunk, NER, GED and Keyword, while the rest are based on the Acc metric~\cite{liu2019linguistic}.
It is worth to note that the vocabulary size of each retrieval dataset differs significantly, which would impact the performance of downstream probing tasks. To make a fair comparison, we remove the instances where the tokens is out of the target vocabulary (i.e., the vocabulary of retrieval datasets) from the probe datasets. All the probing tasks are tuned with Adam with batch size of 80, using a learning rate of 0.001 for maximum number of 50 epochs, using early stopping with a patience of 3.
 
 \begin{table}[!h] 
\begin{tabular}{c c c c c}
\toprule
Probing Tasks & Train & Dev & Test & Metric \\
\hline
    Chunk  & 6k & 1.7k & 1.7k & F1\\
    POS  & 11k & 1.8k & 1.8k & Acc\\
    NER  & 12k & 2.7k & 2.9k & F1\\
    \hline
    GED  & 27k & 2.1k & 2.7k & F1 \\
    SynArcPred & 11k & 1.8k & 1.8k & Acc\\
    SynArcCls & 11k & 1.8k & 1.8k & Acc\\
    Word Scramble & 49k & 8k & 8k & Acc\\
    \hline
    PS-fxn   & 2.4k & 0.5k & 0.5k & Acc \\
    PS-role  & 2.4k & 0.5k & 0.5k & Acc \\
    CorefArcPred & 2.2k& 0.2k&0.2k& Acc\\
    SemArcPred  & 25k & 2.5k & 2.5k & Acc\\
    SemArcCls & 25k & 2.5k & 2.5k & Acc\\
    Synonym  & - & -& 10k & Acc\\
    Polysemy  &- &- & 7.6k & Acc\\
    Keyword Extract  & 109k & 50k & 50k & F1\\
    Topic Classification & 100k & 20k & 20k & Acc\\
    \hline
\bottomrule
\end{tabular}
\caption{Statistics of the datasets of each probing task.}
\label{table:probe_datasets}
\end{table}

\section{References to State-of-the-Art Task-Specific Models (Without Pretraining)}
\label{appendix:prob_results}

\begin{table}[!h] \label{References to State-of-the-Art Models on Retrieval Tasks}
\begin{tabular}{l l}
\toprule
\multirow{2}{*}{Task} & 
\multirow{2}{*}{\tabincell{c}{Previous state of the art\\ (without pretraining)}}
\\ \\\hline
\multirow{1}{*}{POS} &95.82 (\citeauthor{yasunaga2017robust},\citeyear{yasunaga2017robust})  \\
\multirow{1}{*}{Chunk} & 95.77 (\citeauthor{hashimoto2016joint}, \citeyear{hashimoto2016joint})\\
\multirow{1}{*}{NER} & 91.38 (\citeauthor{hashimoto2016joint}, \citeyear{hashimoto2016joint})\\
\multirow{1}{*}{GED} & 39.83 (\citeauthor{rei2019jointly}, \citeyear{rei2019jointly})\\
\multirow{1}{*}{PS-Role} & 66.89 (\citeauthor{schneider2018comprehensive}, \citeyear{schneider2018comprehensive})\\
\multirow{1}{*}{PS-Fxn} & 78.29 (\citeauthor{schneider2018comprehensive}, \citeyear{schneider2018comprehensive})\\
\multirow{1}{*}{Keyword Exaction} & 56.09 (\citeauthor{tixier2016graph}, \citeyear{tixier2016graph})\\
\multirow{1}{*}{Topic Classification} & 76.26 (\citeauthor{wang2018disconnected},\citeyear{wang2018disconnected})\\
\bottomrule
\end{tabular}
\caption{Performance of prior state of the art models (without pretraining) for each probe task.}
\end{table}

\end{document}